\documentclass[twocolumn,prd,showpacs,preprintnumbers,amsmath,amssymb,floatfix]{revtex4}

\usepackage{graphicx}
\usepackage{dcolumn}
\usepackage{bm}

\newcommand{\Sg}{\Sigma}
\newcommand{\Sgs}{\Sigma^*}
\newcommand{\Xs}{\Xi^*}
\newcommand{\bk}{\bar{K}}
\newcommand{\bks}{\bar{K}^*}
\newcommand{\be}{\begin{equation}}
\newcommand{\ee}{\end{equation}}
\newcommand{\ba}{\begin{eqnarray}}
\newcommand{\ea}{\end{eqnarray}}

\newcommand{\Lds}{\Lambda(1520)}
\newcommand{\Ls}{\Lambda^*}
\newcommand{\eps}{\epsilon}
\newcommand{\C}{{\cal{C}}}
\newcommand{\cL}{{\cal{L}}}

\newcommand{\bra}{\langle}
\newcommand{\ket}{\rangle}
\newcommand{\keth}[1]{| \, #1 \, \rangle}

\newcommand{\Tr}{\text{Tr}}

\begin{document}

\preprint{}

\title{Coupling of $\bar{K}^*N$ to the $\Lambda(1520)$}

\author{
T. Hyodo$^{1 *}$\footnotetext{$^*$Electronic address: 
hyodo@rcnp.osaka-u.ac.jp}, 
Sourav~Sarkar$^{2 \dagger}$\footnotetext{$^\dagger$Present address:  
Variable Energy Cyclotron Centre, 1/AF Bidhannagar, Kolkata-700064, India},
A. Hosaka$^{1}$, and
E.~Oset$^2$}

\affiliation{$^1$Research Center for Nuclear Physics (RCNP),
Ibaraki, Osaka 567-0047, Japan. \\
$^2$Departamento de F\'isica Te\'orica and IFIC,
Centro Mixto Universidad de Valencia-CSIC,
Institutos de Investigaci\'on de Paterna, Aptd. 22085, 46071
Valencia, Spain.
}

\date{\today}
\begin{abstract}
    We study the coupling of the $\Lambda(1520)\equiv \Lambda^*$ resonance 
    to the $\bar{K}^*$ vector meson and nucleon. This coupling is not 
    directly measured from the resonance decay, but is expected to be 
    important in hyperon production reactions, in particular for the exotic 
    $\Theta^+$ production. We compute the coupling in two different schemes,
    one in the chiral unitary model where the $\Lambda^*$ is dominated by 
    the quasibound state of mesons and baryons, and the other in the quark 
    model where the resonance is a $p$-wave excitation in the three valence 
    quarks. Although it is possible to construct both models such that they 
    reproduce the $\bar{K}N$ and $\pi\Sigma$ decays, there is a significant 
    difference between the $\Lambda^*\bar{K}^*N$ couplings in the two 
    models. In the chiral unitary model $|g_{\Lambda^*\bar{K}^* N}| \sim 
    1.5$, while in the quark model $|g_{\Lambda^*\bar{K}^* N}| \sim 10$. The
    difference of the results stems from the different structure of the 
    $\Lambda^*$ in both models, and hence, an experimental determination of 
    this coupling would shed light on the nature of the resonance.
\end{abstract}
    

%
\maketitle

\section{Introduction}

Recent activities in hadron physics have been much stimulated by the 
discussions on exotic states. The existence of the exotic pentaquark 
$\Theta^+$~\cite{Nakano:2003qx} is not yet confirmed, but much of the 
works are related to explain its expectedly unusual properties.  

Exotic states, by definition, contain more than three quarks in the case of 
baryons, and more than one quark-antiquark pair in the case of mesons. In 
both cases, the exotic states may have components of two or more color 
singlet states. If the color-singlet correlations such as 
$[\bar{q}q]_{\text{singlet}}$ and $[qqq]_{\text{singlet}}$ are strong, the 
states may be regarded as composite states of two or more hadrons. However, 
if the color-nonsinglet correlations such as diquark correlations are 
strong, the components of color singlet states are only a small part of the 
exotic states.   

Such color-singlet or color-nonsinglet correlations may be tested not only 
in the manifestly exotic states but also in ordinary hadrons. The role of 
diquark correlations in hadrons has been discussed~\cite{Jaffe:2004ph,
Jaffe:2005md}. Contrary, the importance of color-singlet correlations may be
tested by the mesonic cloud around baryons. For instance, the existence of a
pion cloud offers an explanation of the negative charge radius of the 
neutron. The strong correlation between mesons and baryons, as implied by 
chiral perturbation theory, has been shown to generate baryon resonances 
especially in $s$-wave scattering channels: For instance, the 
$\Lambda(1405)$ resonance which can be generated in $s$-wave $\bar{K}N$ 
scattering~\cite{annphys10.307,Jennings:1986yg,Kaiser:1995eg,Oset:1998it}. 
An interesting feature of such a dynamically generated $\Lambda(1405)$ is 
that it is a superposition of two poles near the nominal mass region, one of
which couples dominantly to the $\bar{K}N$ and the other to $\pi \Sigma$ 
state~\cite{Jido:2003cb,Hyodo:2003jw,Magas:2005vu}.

Recently, another $\Lambda$ resonance, the $\Lambda(1520)\equiv \Lambda^*$ 
of $J^P = 3/2^-$, has been investigated in several contexts. In 
Refs.~\cite{Kolomeitsev:2003kt,Sarkar:2004jh}, the resonance was described 
as a quasibound state of $\pi \Sigma(1385)$ and $K \Xi(1530)$ in $s$ wave. 
In these studies, the identification of some baryon resonances with $s$-wave
quasibound state of an octet meson and a decuplet baryon has been 
extensively studied. This approach is further extended in particular to the 
$\Lambda^*$, by including the $d$-wave channels of mesons and ground state 
baryons~\cite{Sarkar:2005ap,Luis,Sarkar:2005sp}, leading to a successful 
description of existing data.

The $\Lambda^*\bar{K}^*N$ coupling is worth being studied. In the 
experimental data~\cite{Barber:1980zv} and its analysis for $\Lambda^*$ 
photoproduction~\cite{Sibirtsev:2005ns}, the important role of $\bar{K}^*$ 
vector meson was suggested, while a similar behavior was recently explained 
by means of the photo-$K^*$ contact term~\cite{Nam:2005uq}. Not much is 
known for the properties of the interaction with $\bar{K}^*$, which is 
expected to be important in associated $\Lambda^*$ and $\Theta^+$ production
from deuteron as observed recently by the LEPS 
collaboration~\cite{Titov:2005kf}.
As compared to the interactions with a kaon, we must rely much on models for
the estimation of the $\bar{K}^*$ interaction, since there is no theoretical
framework to introduce it such as chiral symmetry, nor experimental 
information on the decay of the $\Lambda^*$ to $\bar{K}^* N$, which is 
kinematically forbidden.

In this paper, we investigate exclusively the $\bar{K}^*$ coupling to the
$\Lambda^*$, where the $\Lambda^*$ is formed dominantly by the $s$-wave
$\pi\Sigma(1385)$ quasibound state, which is supplemented by the 
$K\Xi(1530)$ state and the $d$-wave $\bar{K} N$ and $\pi \Sigma$ states. 
Since this is the first attempt to investigate the quantity in the present 
framework, we explain in detail how we compute the coupling in the present 
model. The result is then compared with that of the conventional quark 
model, where the $\Lambda^*$ is described as a $p$-wave excitation of one 
of the three valence quarks. This comparison should be useful in testing 
the very different nature of the two descriptions, as we will discuss in 
detail. 

This paper is organized as follows. In Sec.~\ref{sec:formulation}, we 
describe how the $\Lambda^*\bar{K}^*N$ coupling is computed in the chiral 
unitary model for $\Lambda^*$. Numerical results and discussions are 
presented in Sec.~\ref{sec:result}, where we compare the result of the 
chiral unitary model with the quark model predictions. The final section is 
devoted to summarize the present work.  

\section{Formulation}\label{sec:formulation}

\subsection{Structure of the amplitude}


We consider an effective interaction Lagrangian~\cite{Nam:2005uq} given by
\begin{equation}
   \mathcal{L}_{\Lambda^*\bar{K}^*N}
   =\frac{g_{\Lambda^*\bar{K}^*N}}{M_{K^*}}
   \bar{\Lambda}^*_{\mu}\gamma_{\nu}
   (\partial^{\mu}K^{*\nu}-\partial^{\nu}K^{*\mu})N
   +\text{h.c.}~,
   \label{eq:Lagrangian}
\end{equation}
where $M_{K^*}$ is the mass of the vector $K^*$ meson, h.c. denotes the 
hermitian conjugate, and $g_{\Lambda^*\bar{K}^*N}$ is the coupling constant.
Because $J^P(\Lambda^*)=3/2^-$, the coupling has two independent components.
In terms of multipoles, they are $E1$ and $M2$, which are related to the two
helicity amplitudes $A_{1/2}$ and $A_{3/2}$. In the $E1$ amplitude, the 
orbital angular momentum of the decaying channel of $\bar{K}^*N$ is $s$ 
wave, while in $M2$, it is $d$ wave. Here, we investigate the $s$-wave 
coupling which is the $E1$ amplitude in the chiral unitary model. We expect 
that the $s$-wave coupling dominates in the small three-momentum $|\bm{k}|$ 
region, where $\bm{k}$ is the relative momentum of the (virtual) $\bar{K}^*$
and $N$. Assuming the interaction region of about 1 fm, the $d$-wave and 
hence the $M2$ component will become important for $|\bm{k}| > 400$ MeV. 

Applying the nonrelativistic reduction to Eq.~\eqref{eq:Lagrangian}, and 
picking up the $s$-wave component, we obtain the transition amplitude of 
$\bar{K}^*N\to\Lambda^*$ as
\begin{equation}
    -it_{\Lambda^*\bar{K}^*N}
    = g_{\Lambda^*\bar{K}^*N}\bm{S}^\dagger\cdot\bm{\eps}.
    \label{eq:coupling}
\end{equation}
Here $\bm{\epsilon}$ is the polarization vector of the $\bar{K}^*$ and
$\bm{S}$ is the spin transition operator~\cite{Ericson:1988gk}, which is 
defined by $ \bra 3/2, m+\lambda | S^{\dag}_\lambda | 1/2, m\ket = 
\C(\tfrac{1}{2}\ 1\ \tfrac{3}{2};m,\lambda)$ where $\lambda$ represents a 
spherical component $\pm 1$ or 0 and $\C(j_1\ j_2\ J;\mu_1,\mu_2)$ denotes 
the SU(2) Clebsch-Gordan coefficient for $\bm{J}(\mu_1+\mu_2)=
\bm{j}_1(\mu_1)+\bm{j}_2(\mu_2)$.

\begin{figure}[tbp]
\centerline{\includegraphics[width=6cm]{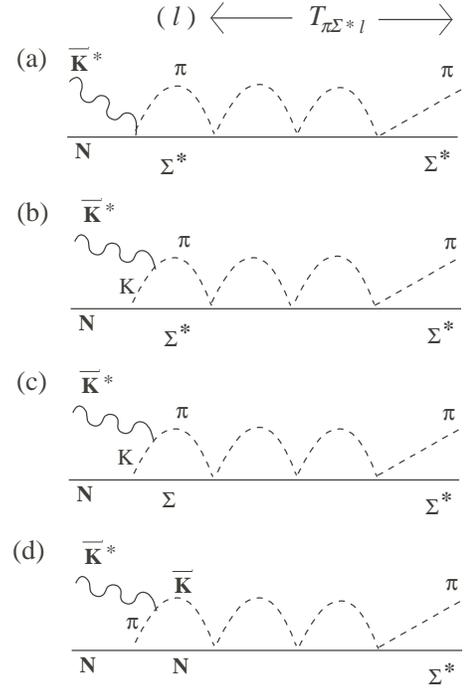}}
\caption{Diagrams for the microscopic mechanism of $\bar{K}^*N\to 
\Lambda^*\to \pi\Sigma^*$ calculated in the chiral unitary model.}
\label{fig:mech1}
\end{figure}

In the chiral unitary model, the $\Lambda^*$ is generated dynamically in the
scattering of the $\pi\Sgs$ and $K\Xs$ channels in $s$ wave and the $\bk N$ 
and $\pi\Sg$ channels in $d$ wave~\cite{Sarkar:2005ap,Luis}. In order to 
estimate the coupling of the $\Lambda^*$ resonance to the $\bks N$ channel, 
we follow the microscopic mechanism as illustrated in Fig.~\ref{fig:mech1}.
In this case, the $\bks N$ couples to the dynamically generated $\Lambda^*$,
represented by the amplitude $T$ in the figure, decaying into the 
$\pi\Sigma^*$ channel. Notice that the $K\Xi^*$ channel does not appear in 
the first intermediate loop, since there is no direct coupling from 
$\bar{K}^*N$ to $K\Xi^*$. Schematically, the process $\bar K^* N \to 
\Lambda^* \to \pi \Sigma(1385)$ can be expressed as
\begin{equation}
    -it_{\text{ChU}}
    =\sum_{l} (-iT_{\pi\Sigma^* l}) iG_{l} (-it_{l\bar{K}^*N}),
    \label{eq:tChU}
\end{equation}
where $T_{\pi\Sigma^* l}$ is $l\to \pi\Sigma^*$ amplitude obtained by the 
chiral unitary model~\cite{Sarkar:2005ap,Luis}, $G_l$ is the loop function 
of the intermediate state $l$, and $-it_{l\bar{K}^*N}$ is the amplitude of 
$\bar{K}^*N\to l$. As shown in Fig.~\ref{fig:mech1}, there are four types of
transition amplitudes for $-it_{l\bar{K}^*N}$ with three different 
intermediate states $\pi \Sigma^*, \pi \Sigma$, and $\bar K N$. 

\begin{figure}[tbp]
\centerline{\includegraphics[width=6cm]{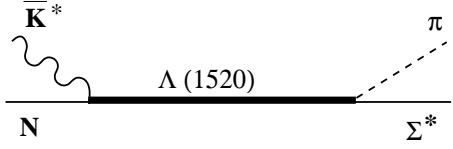}}
\caption{Diagram for the resonance dominance model of $\bar{K}^*N\to 
\Lambda^*\to \pi\Sigma^*$.}
\label{fig:mech2}
\end{figure}

Since we are considering first the $s$-wave coupling, the amplitude 
$-it_{l\bar{K}^*N}$ should be written as $-it_{l\bar{K}^*N}=g_{l\bar{K}^*N}
\bm{S}^\dagger\cdot\bm{\eps}$, where $g_{l\bar{K}^*N}$ will be calculated 
later. We denote the total energy as $\sqrt{s}$, and consider the energy 
region close to the $\Lambda^*$ pole $\sqrt{s}\sim M_{\Ls}$ with 
$M_{\Lambda^*}$ being the mass of the $\Ls$ resonance. In this region, 
the chiral unitary amplitude $T_{ij}$ can be approximated by the 
Breit-Wigner propagator $T_{ij}\sim g_{\Lambda^*i}g_{\Lambda^*j}/
(\sqrt{s}-M_{\Lambda^*})$ with coupling constants $g_{\Lambda^*i}$, where
$i$ stands for the channels coupling to $\Ls$. Then we have
\begin{equation}
    -it_{\text{ChU}}
    \sim -ig_{\Lambda^*\pi\Sigma^*} \frac{i}{\sqrt{s}-M_{\Lambda^*}}
    \sum_{l}g_{\Lambda^* l} \, G_{l} \, g_{l\bar{K}^*N}
    \bm{S}^\dagger\cdot\bm{\eps}.
    \label{eq:ChUamp}
\end{equation}
On the other hand, with the $s$-wave coupling Eq.~\eqref{eq:coupling}, the 
resonance model for the amplitude $\bar K^* N \to \Ls \to \pi \Sigma(1385)$
can be written as shown in Fig.~\ref{fig:mech2}, 
\begin{equation}
    -it_{\text{res}}=-ig_{\Ls\pi\Sgs}\frac{i}{\sqrt{s}-M_{\Ls}}
    g_{\Ls\bks N}\bm{S}^\dagger\cdot\bm{\eps},
    \nonumber
\end{equation}
where $g_{\Ls\bks N}$ is the $\Ls\bks N$ coupling constant that we are 
interested in. Hence comparing this amplitude with Eq.~\eqref{eq:ChUamp},
we extract the $\Ls\bks N$ coupling as
\begin{equation}
    g_{\Ls\bks N}
    =\sum_{l}g_{\Lambda^* l}\, G_{l}\, g_{l\bar{K}^*N}.
    \label{eq:couplingformula}
\end{equation}
In the previous study~\cite{Luis}, the coupling constants $g_{\Lambda^* l}$ 
have been determined as
\begin{equation}
    g_{\Ls\pi\Sgs}=0.91,
    \quad
    g_{\Ls\pi\Sg}=-0.45,
    \quad
    g_{\Ls\bk N}=-0.54,
    \label{eq:gLambda}
\end{equation}
which well reproduce the partial decay widths of the $\Lambda(1520)$ 
resonance to these channels. In the following, we evaluate
$G_{l}g_{l\bar{K}^*N}$ by calculating the diagrams in Fig.~\ref{fig:mech1}
one by one.

\subsection{Computation of loop diagrams}

\begin{figure}[tbp]
\centerline{\includegraphics[width=6cm]{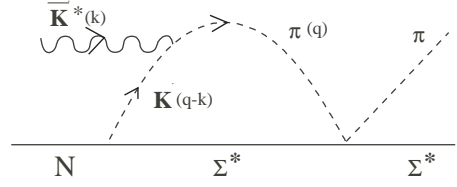}}
\caption{Momentum assignment for the diagram (b) in Fig.~\ref{fig:mech1}.}
\label{fig:piSstarloop}
\end{figure}

Let us first consider the diagrams (a) and (b) in Fig.~\ref{fig:mech1}. The 
amplitudes for these diagrams $-it^{(a)}$ and $-it^{(b)}$ are related to 
each other through the gauge condition
\begin{equation}
    (-it_{\mu}^{(a)}-it_{\mu}^{(b)})k^{\mu}=0,
    \label{eq:gauge}
\end{equation}
where $-it^{(i)}\equiv -it^{(i)}_{\mu}\epsilon^{\mu}$ and $k^{\mu}$ is the 
momentum of the $\bar{K}^*$. First we consider the diagram (b). Utilizing 
the interaction Lagrangians given in appendix, the amplitude of, for 
instance, $\bar{K}^{*0}n\to \pi^+\Sigma^{*-}$ for the meson pole diagram (b)
at tree level is written as
\begin{align}
    -it^{(b)}_{\pi^-\Sigma^{*+}\bar{K}^{*-}n}
    =&\frac{1}{\sqrt{2}}ig\epsilon^{\mu}(2q_{\mu}-k_{\mu})
    \frac{i}{(q-k)^2-m_K^2} \nonumber \\
    &\times \frac{1}{\sqrt{3}}
    \frac{g_A^*}{2f}\bm{S}^{\dag}\cdot (\bm{q}-\bm{k}).
    \label{eq:amp_b1}
\end{align}
The momentum variables in Eq.~\eqref{eq:amp_b1} are assigned as shown in 
Fig.~\ref{fig:piSstarloop}, $m_K$ is the mass of kaon, $g=-6.05$, 
$g_A^*=(D+F)\times 2.13$, $D+F=1.26$, $f=f_\pi=93$ MeV. In order to obtain 
the corresponding tree level amplitude for the contact diagram (a), 
$-it^{(a)}_{\pi^-\Sigma^{*+}\bar{K}^{*-}n}$, we first replace 
$\epsilon^{\mu}$ by $k^{\mu}$ in Eq.~\eqref{eq:amp_b1}, set $q^2=m_{\pi}^2=
M_K^2$ assuming the SU(3) limit (this manipulation is only for the purpose 
of determining the contact term) and set $\bm{q}=\bm{0}$. Then, the contact 
term has to be 
\begin{equation}
    -it^{(a)}_{\pi^-\Sigma^{*+}\bar{K}^{*-}n}
    =\frac{g}{\sqrt{2}}\frac{g_A^*}{2f}\frac{1}{\sqrt{3}}
    \bm{S}^{\dag}\cdot \bm{\epsilon},
    \nonumber
\end{equation}
in order to satisfy Eq.~\eqref{eq:gauge}. 

We can repeat the same operation for other charge states. Writing the 
$\bar{K}^*N$ and $\pi\Sigma^*$ states in isospin basis (recalling that 
$\keth{K^{*-}} = -\keth{1/2,-1/2}$ and $\keth{\pi^+} = -\keth{1,1}$ in our 
convention), we find
\begin{equation}
    -it^{(a)}_{\pi\Sigma^{*}\bar{K}^{*}N}
    =\frac{g}{2}\frac{g_A^*}{2f}
    \bm{S}^{\dag}\cdot \bm{\epsilon},
    \label{eq:amp_a1}
\end{equation}
after projecting over $I=0$. Inserting Eq.~\eqref{eq:amp_a1} into 
Eq.~\eqref{eq:tChU}, we can now write
\begin{equation}
    -it^{(a)}= 
    (-iT_{\pi\Sgs\pi\Sgs}) i G_{\pi\Sgs} \, 
    g_{\pi\Sigma^*\bar{K}^*N} \,
    \bm{S}^\dagger\cdot\bm{\eps},
    \label{eq:amp_a}
\end{equation}
where $G_{\pi\Sgs}$ is the loop function involving the $\pi$ and the $\Sgs$:
\begin{align}
    G_{\pi\Sgs}(\sqrt{s})
    =& i\int \frac{d^4q}{(2\pi)^4}
    \frac{1}{q^2-m_{\pi}^2+i\epsilon} \nonumber \\
    &\times\frac{1}{\sqrt{s}-q^0-E_{\Sigma^*}+i\epsilon},
    \nonumber
\end{align}
where $E_{\Sgs}(\bm{q})=\sqrt{M_{\Sgs}^2+\bm{q}^2}$ and the coupling 
constant is given by
\begin{equation}
    g_{\pi\Sigma^*\bar{K}^*N}
    =\frac{1}{2}g\frac{g_A^*}{2f}.
    \label{eq:1a}
\end{equation}

On the other hand, we can also extract the $s$-wave component of the meson 
pole term from Eq.~\eqref{eq:amp_b1} after projecting over $I=0$, and we 
find
\begin{equation}
    -it^{(b)}_{l\bar{K}^*N}
    =g_{\pi\Sigma^*\bar{K}^*N}
    \frac{2}{3}\frac{\bm{q}^2}{(q-k)^2-m_K^2}
    \bm{S}^{\dag}\cdot \bm{\epsilon},
    \label{eq:amp_b2}
\end{equation}
where the variable $\bm{q}$ should be included in the loop function. 
Therefore, the amplitude for this process can be expressed similarly as in 
Eq.~\eqref{eq:amp_a} but with the meson-baryon loop function $G_{\pi\Sgs}$ 
replaced by the loop function with an additional factor, which is defined by
\begin{align}
    \tilde{G}_{\pi\Sgs K}(\sqrt{s},k)
    =&i\int \frac{d^4q}{(2\pi)^4}
    \frac{\bm{q}^2}{(q-k)^2-m_K^2+i\epsilon} 
     \nonumber \\
    &\times \frac{1}{q^2-m_{\pi}^2+i\epsilon}
    \frac{1}{\sqrt{s}-q^0-E_{\Sigma^*}+i\epsilon} .
    \nonumber
\end{align}
Finally, combining the contributions from $(a)$ and $(b)$, we obtain
\begin{align*}
    -it^{(a)}-it^{(b)}
    =& (-iT_{\pi\Sgs\pi\Sgs}) 
    i\left(
    G_{\pi\Sgs} +\frac{2}{3}\tilde G_{\pi\Sgs K}
    \right)
    \nonumber \\
    &\times
    g_{\pi\Sigma^*\bar{K}^*N}
    \bm{S}^\dagger\cdot\bm{\eps}~.
\end{align*}

\begin{figure}
\centerline{\includegraphics[width=6cm]{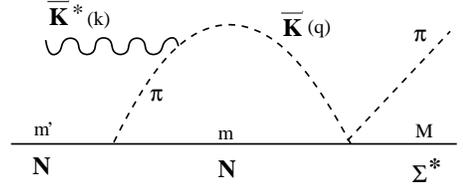}}
\caption{Momentum and spin indices assignment for the loop diagram
in (d) in Fig.~\ref{fig:mech1}.}
\label{fig:kbarnloop}
\end{figure}

We now evaluate the amplitude for the diagram (c) and (d) in 
Fig.~\ref{fig:mech1}. The structure of the first loop can be found from 
Fig.~\ref{fig:kbarnloop}. Since we need the $d$-wave projection of the meson
pole term to balance the $d$-wave $\bar{K}N\to \pi\Sigma^*$ amplitude in the
loop, we study the amplitude in some detail. Using the interaction 
Lagrangians given in the appendix, the $I=0$ component of the tree level 
amplitude for (d), for instance, is given by
\begin{align*}
    -it^{(d)}_{\bar{K}N\bar{K}^*N}
    =&-3ig\frac{D+F}{2f}\eps^\mu(2q_{\mu}-k_{\mu}) \\
    &\times\frac{i}{(q-k)^2-m_{\pi}^2}
    \bm{\sigma}\cdot(\bm{k}-\bm{q}).
\end{align*}
The spin structure takes the form $(\bm{\eps}\cdot\bm{q})
(\bm{\sigma}\cdot \bm{q})$, neglecting $\bm{k}$ which is assumed to be 
small. Now, the $d$-wave structure obtained from $\sigma_iq_i\eps_jq_j\to
\sigma_i\eps_j(q_iq_j-\bm{q}^2\delta_{ij}/3)$ will combine with the 
$d$-wave structure coming from the $\bk N\to\pi\Sgs$ vertex to produce a 
scalar quantity after the loop integration is performed. We write
\begin{equation}
    \sigma_i\eps_j(q_iq_j-\tfrac{1}{3}\bm{q}^2\delta_{ij})
    =A\left[[\sigma\otimes\eps]_\mu^2\ Y_2(\hat q)\right]_0^0,
    \label{eq:dwave}
\end{equation}
where $A$ is a constant. This indicates that the two vector operators 
$\sigma$ and $\eps$ combine to produce an operator of rank $2$, which 
couples to the spherical harmonic $Y_2(\hat q)$ to produce a scalar. The 
right hand side can be written as
\begin{align*}
    A&\sum_\mu(-1)^\mu [\sigma\otimes\eps]_\mu^2\
    Y_{2,-\mu}(\hat q) \\
    &=
    A\sum_{\mu,\alpha}(-1)^\mu Y_{2,-\mu}(\hat q)\ 
    \C(1\ 1\ 2;\alpha,\mu-\alpha)\sigma_\alpha\eps_{\mu-\alpha}.
\end{align*}
To find the value of $A$ we take the matrix element 
of both sides of Eq.~\eqref{eq:dwave} between the states 
$m$ and $m'$ so that
\begin{align}
  \langle m|\sigma_i &\eps_j
  \left(q_iq_j-\frac{1}{3}
  |\bm{q}|^2\delta_{ij}\right)|m'\rangle 
  \nonumber \\
  =&
  A\sum_\mu(-1)^\mu\ Y_{2,-\mu}(\hat q)\ \eps_{\mu-m+m'}
  \nonumber \\
  &\times \C(1\ 1\ 2;m-m',\mu-m+m')
  \nonumber \\
   &\times
   \C\left(\frac{1}{2}\ 1\
  \frac{1}{2};m',m-m'\right),
\label{eq:dwave1}
\end{align}
where we have used $\langle m|\sigma_\alpha|m'\rangle=\sqrt{3}\ 
\C(\tfrac{1}{2}\ 1\ \tfrac{1}{2};m',\alpha)$ with $m=m^{\prime}+\alpha$. 
Considering specific values of $m$ and $m'$, we obtain
\begin{equation}
    A=\sqrt{\frac{8\pi}{15}}\ \bm{q}^2.
    \label{eq:a}
\end{equation}

Following Ref.~\cite{Sarkar:2005ap}, we now include the $\bk N\to\pi\Sgs$ 
vertex given by
\begin{align}
    -it_{\bk N\to\pi\Sgs} 
    =&-i\beta_{\bk N}\ \bm{q}^2\ \C\left(\frac{1}{2}\ 2\
    \frac{3}{2};m,M-m\right)
    \nonumber \\
    &\times Y_{2,m-M}(\hat{q})(-1)^{M-m}\sqrt{4\pi},
    \label{eq:Kbarnamp}
\end{align}
so that the total spin structure of the loop shown in 
Fig.~\ref{fig:kbarnloop} is essentially given by
\begin{align*}
    J=&\sum_m\int\frac{d\Omega_q}{4\pi}
    \langle m|\sigma_i\eps_j
    \left(q_iq_j
    -\frac{1}{3}\bm{q}^2\delta_{ij}\right)|m'\rangle
    \\
    & \times \C\left(\frac{1}{2}\ 2\ \frac{3}{2};m,M-m\right)
    Y_{2,m-M}(\hat{q})(-1)^{M-m}\sqrt{4\pi},
\end{align*}
where we perform an average over the angular dependence in the integration 
over the loop momentum $q$. Using Eqs.~\eqref{eq:dwave1} and \eqref{eq:a} 
this can be written as
\begin{align*}
    J=&\sqrt{\frac{2}{3}} \bm{q}^2 (-1)^{1-M+m'}\ \eps_{m'-M}
    \sum_m\ \C(\tfrac{1}{2}\ 1\ \tfrac{1}{2};m',m-m')\ 
    \\
    &\times
    \C(\tfrac{1}{2}\ 2\ \tfrac{3}{2};m,M-m)\C(1\ 2\ 1;m-m',M-m),
\end{align*}
where we have used the well known relations
\begin{equation}
    \int\ d\Omega_q\ Y_{2,-\mu}(\hat q)\ 
    Y_{2,m-M}(\hat q)=(-1)^\mu\delta_{\mu,m-M},
    \nonumber
\end{equation}
and
\begin{align*}
    &\C(1\ 1\ 2;m-m',m'-M) \\
    =&(-1)^{1-m+m'}\sqrt{\frac{5}{3}}\ 
    \C(1\ 2\ 1;m-m',M-m).
\end{align*}
The product of three Clebsch-Gordan coefficients is then combined into a 
single one with Racah coefficients, resulting in the identity
\begin{align}
     \sum_m &
     \C\left(\frac{1}{2}\ 1\ \frac{1}{2};m',m-m'\right)\
     \C\left(\frac{1}{2}\ 2\
     \frac{3}{2};m,M-m\right) \nonumber \\
     &\times \C(1\ 2\ 1;m-m',M-m)\nonumber\\
     &=-\sqrt{\frac{1}{2}}\ \C\left(\frac{1}{2}\ 1\ 
     \frac{3}{2};m',M-m'\right),
     \nonumber
\end{align}
so that, we finally have
\begin{equation}
    J=\frac{1}{\sqrt{3}}\ \bm{q}^2\ \bm{S}^\dagger\cdot\bm{\eps}.
    \label{eq:J}
\end{equation}
The above relation implies that for practical purposes we can replace in the
first vertex $(\bm{\epsilon}\cdot\bm{q})(\bm{\sigma}\cdot\bm{q})$ by the 
simple form $\tfrac{1}{\sqrt{3}}\ \bm{q}^2\ \bm{S}^\dagger\cdot\bm{\eps}$ 
and for the second vertex the factor $\beta_{\bk N}\bm{q}^2$ and continue
with the formalism exactly as in $s$-wave. Putting everything together, the 
amplitude for the process shown in Fig.~\ref{fig:mech1} (d) can be written
as
\begin{equation}
    -it^{(d)}=(-iT_{\pi\Sgs\bk N })
    i\tilde G_{\bk N\pi}\, 
    g_{\bk N\bks N} \bm{S}^\dagger\cdot\bm{\eps},
    \label{eq:finalkbarn}
\end{equation}
which has the same form as Eq.~\eqref{eq:1a}. In the above equation, we have
defined
\begin{equation}
    g_{\bk N\bks N}=\sqrt{3}g\frac{D+F}{2f},
    \nonumber
\end{equation}
and 
\begin{align}
    \tilde G_{\bk N\pi}(\sqrt{s},k) 
    =&i\int\frac{d^4q}{(2\pi)^4}\
    \frac{\bm{q}^2}{(q-k)^2-m_\pi^2+i\eps}\
    \frac{\bm{q}^2}{q_{\text{on}}^2}\nonumber\\
    &\times \frac{1}{q^2-m_K^2+i\eps} \frac{M_N}{E_N}
    \frac{1}{\sqrt{s}-q^0-E_N+i\eps},
    \label{eq:gtkbarn}
\end{align}
with $q_{\text{on}}=\lambda^{1/2}(s,m_N^2,m_K^2)/2\sqrt{s}$. The factor 
$\bm{q}^2$ appearing in the vertex of Eq.~\eqref{eq:J} is kept in the 
loop. On the other hand, the amplitudes which we use for $\bar{K}N\to
\pi\Sigma^*$ of Eq.~\eqref{eq:finalkbarn} factorize the on shell value 
$q_{\text{on}}^2$. This is the reason for the factor $\frac{\bm{q}^2}
{q_{\text{on}}^2}$ in Eq.~\eqref{eq:gtkbarn} since in 
Eq.~\eqref{eq:finalkbarn} we write explicitly $T_{\pi\Sgs\bk N }$.

The amplitude for the process shown in Fig.~\ref{fig:mech1} (c) can be 
evaluated in a similar way as described above. In this case we have
\begin{equation}
    -it^{(c)}=(-iT_{\pi\Sgs\pi\Sg}) i\tilde G_{\pi \Sg K}
    g_{\pi \Sg \bks N} \bm{S}^\dagger\cdot\bm{\eps},
    \nonumber
\end{equation}
where
\begin{equation}
    g_{\pi \Sg\bks N}=\sqrt{2}g\frac{D-F}{2f},
    \nonumber
\end{equation}
with $D-F=0.33$ and $G_{\pi \Sg K}$ given similarly as in 
Eq.~(\ref{eq:gtkbarn}) with the replacements $\pi\to K$ and $N\to \Sg$.


Following Eq.~\eqref{eq:couplingformula}, we thus obtain the coupling of the
$\Lds$ with $\bks N$ as
\begin{align}
    g_{\Ls\bks N}(\sqrt{s},k) 
    =&  g_{\Ls\pi\Sgs} \left[G_{\pi\Sgs}(\sqrt{s})
    +\frac{2}{3}\tilde G_{\pi\Sgs K}(\sqrt{s},k)\right]\nonumber\\
    &\times g_{\pi\Sgs \bks N}  + g_{\Ls\pi\Sg}\,
    \tilde G_{\pi\Sg K}(\sqrt{s},k)\, g_{\pi\Sg\bks N}
    \nonumber \\
    &+ g_{\Ls\bk N} \, \tilde G_{\bk N\pi}(\sqrt{s},k) \,
    g_{\bk N\bks N}.
    \label{eq:result}
\end{align}

\section{Results and discussions}\label{sec:result}

\subsection{Chiral unitary model}

Before calculating Eq.~\eqref{eq:result}, let us consider the momentum 
variables. Since Eq.~\eqref{eq:ChUamp} is valid close to the pole of the 
resonance, we choose $\sqrt{s}=1520$ MeV. For this $\sqrt{s}$, $\Lambda^*$ 
cannot decay into $\bar{K}^*(892)$ and $N(940)$. Here we assume that the
$\bar{K}^*$ is off the mass shell with the nucleon being on-shell, which 
would be compatible with the $K^*$ $t$-channel exchange in $\Lambda^*$ 
photoproduction on the nucleon target. Then the energy of the $\bar{K}^*$ 
can be given by
\begin{equation}
    k^0 = \sqrt{s}-E_N(\bm{k}) = \sqrt{s}- \sqrt{M_N^2+\bm{k}^2},
    \nonumber
\end{equation}
where we are in the center of mass frame. As we have seen, our formulation 
is consistent with $|\bm{k}|\sim 0$, where the $s$-wave interaction is 
dominant. If $|\bm{k}|=0$, we obtain $ k^0  =\sqrt{s}-M_N \sim 580$ MeV,
which is the maximum energy of the $\bar{K}^*$ when the nucleon is on-shell.

\begin{figure}[tbp]
    \centering
    \includegraphics[width=8.5cm,clip]{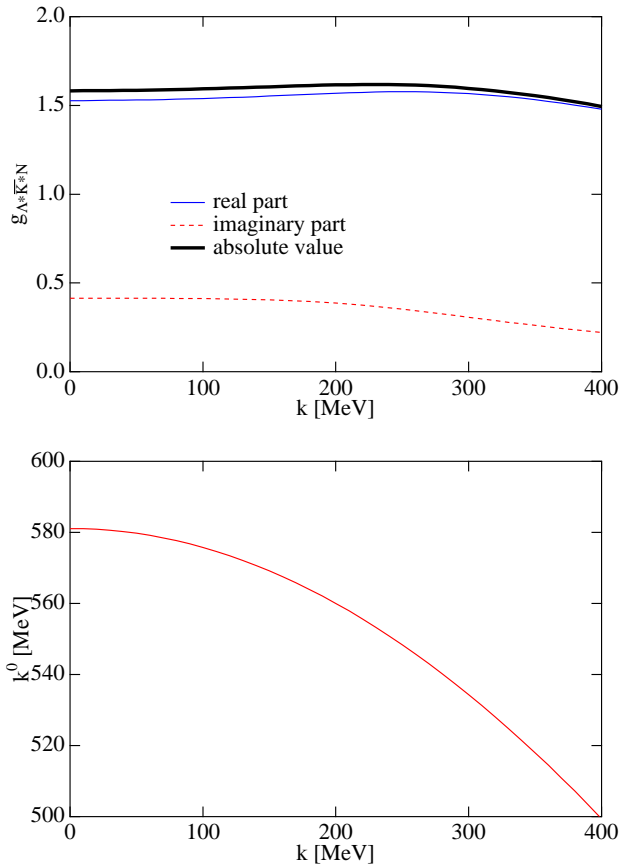}
    \caption{\label{fig:result}
    (Color online) Upper: Numerical results for the $\Lambda^*\bar{K}^*N$
    coupling constant as a function of $K^*$ momentum 
    $|\bm{k}|$ in the chiral unitary model. Thick solid line, thin solid 
    line, and dashed line represent absolute value, real part, and imaginary
    part of the coupling constant, respectively. Lower: Energy of the $K^*$
    as a function of $|\bm{k}|$, assuming the nucleon is on-shell.}
\end{figure}

In order to study the finite momentum effect and stability of the result, 
we vary the momentum $|\bm{k}|$ from zero to 400 MeV, and plot the real and 
imaginary parts as well as the absolute value of the $\Lambda^*\bar{K}^*N$ 
coupling constant in Fig.~\ref{fig:result}. For reference, we also plot the 
energy $k^0$ in the lower panel in Fig.~\ref{fig:result}. We observe that 
the result is stable against the momentum $|\bm{k}|$ up to $\sim 200$ 
MeV, where the $s$-wave coupling is expected to be dominant. Numerical 
values are 
\begin{equation}
    g_{\Lambda^*\bar{K}^*N} \sim 1.53 + 0.41i,\quad
    |g_{\Lambda^*\bar{K}^*N}| \sim 1.58.
    \nonumber
\end{equation}
The complex phase is the relative one to $g_{\Lambda^*\bar{K}N}=-0.45$ 
given in Eq.~\eqref{eq:gLambda}.

Let us look at each component in detail. Substituting the numerical factors,
Eq.~\eqref{eq:result} can be written as
\begin{align}
    g_{\Lambda^*\bar{K}^*N}
    \sim &-0.042
    G_{\pi\Sgs}
    -0.028 \tilde G_{\pi\Sgs K}\nonumber  \\
    &+
    0.0068 \tilde G_{\pi\Sg K}
    +
    0.038 \tilde G_{\bk N\pi}.
    \label{eq:result2}
\end{align}
Note that the contribution from $\tilde G_{\pi\Sg K}$ is factor 5 smaller 
than the others, due to the $D-F$ factor. 

\subsection{Quark model}

In the quark model, $\Lambda(1520)$ resonance is a $p$-wave state of 
70-dimensional representation of SU(6)~\cite{Isgur:1978xj}. In the 
spin-flavor group, it is a superposition of $^2\bm{1}$, $^2\bm{8}$, and 
$^4\bm{8}$. Here we use the notation $^{2S+1}D$, where $2S+1$ is the 
degeneracy of spin states and $D$ denotes a flavor representation. In the 
standard quark model, $\Lambda^*$ is dominated by the flavor singlet 
$^2\bm{1}$ with some mixture of $^2\bm{8}$; the spin quartet $^4\bm{8}$ has 
only a small fraction.  

Such a wave function has been tested for the decay of $\Lambda^* \to 
\bar{K}N, \pi \Sigma$, and has been proven to work reasonably 
well~\cite{Isgur:1978xj,Hey:1974nc}. For the decay to the chiral mesons, 
the matrix elements of the meson-quark interaction can be taken
\begin{align}
     \mathcal{L}_{mqq}
     &= -i g_{mqq} \bar q \gamma_5 \Phi q
     \to \tfrac{g_{mqq}}{2m_q} 
     \chi^\dagger \bm{\sigma} \cdot \bm{\nabla} \Phi \chi~,
     \nonumber 
\end{align}
where $\chi$ is a two-component spinor, and in the second line the 
non-relativistic approximation is performed. The SU(3) meson field is
defined here by 
\ba
\Phi = 
\left(
\begin{array}{c c c}
\pi^0 + \frac{\eta}{\sqrt{3}} & \sqrt{2} \pi^+ & \sqrt{2} K^+ \\
\sqrt{2} \pi^- & \pi^0 - \frac{\eta}{\sqrt{3}} & \sqrt{2} K^0 \\
\sqrt{2}K^- & \sqrt{2}K^0 & - \frac{2\eta}{\sqrt{3}}
\end{array}
\right) \, .
\nonumber
\ea
The meson-quark coupling constant $g_{mqq} \sim 2.6$ is determined from the 
$\pi NN$ coupling $g_{\pi NN} \sim 13$, and the constituent quark mass is 
taken as $330$ MeV for all $u, d, s$ quarks for simplicity. The use of a 
larger mass for $m_s$ will change slightly the SU(6) symmetric wave function
such that the excitation of the strange quark will be easier than the 
excitation of the $u, d$ quarks. But we expect that the following results 
are not affected too much.  

For the $\bar{K}^*$ (vector meson) coupling, we can use the interaction 
Lagrangian at the quark level
\begin{align}
    \mathcal{L}_{vqq} 
    =& 
    g_{vqq} \bar q \gamma_\mu V^\mu q \nonumber \\
    \to & - \frac{g_{vqq}}{\sqrt{2}m_q}
    \Bigl\{
    u^\dagger (i \bm{\nabla} - i \bm{\nabla}) s\nonumber \\
    &+ \bm{\nabla} \times (u^\dagger \bm{\sigma} s) \Bigr\}
    \cdot \bm{\epsilon}(K^{*+}) + \text{h.c.} \, , 
    \label{Lvqq}
\end{align}
where $\bm{\epsilon}(K^{*+})$ is the polarization vector of the $K^{*+}$,
the quark flavor is indicated explicitly for the $\bar{K}^*$ coupling, and 
the $g_{vqq} \sim 3$ is determined by the empirical $\rho NN$ coupling 
strength. This Lagrangian of vector type coupling works well for baryon 
magnetic moments when the $\bar{K}^*$ is replaced by the photon after SU(3) 
rotation. For the $\rho NN$, however, the tensor coupling is slightly 
underestimated $g_T/g_V \sim 4$, as compared with the strong tensor coupling
$g_T/g_V \sim 6$~\cite{Machleidt:1987hj}. For the present study of 
qualitative analysis, however, we simply adopt the Lagrangian~\eqref{Lvqq}.

In order to extract the relevant coupling strength, we compute the two 
transverse helicity amplitudes, 
\begin{align}
    a_{3/2} &\equiv - \bra N(s_z = 1/2), K^*(h = +1) | 
    {\cal L}_{vqq} | \Lambda^* (s_z = 3/2)\ket 
    \nonumber \\
    a_{1/2} &\equiv - \bra N(s_z = -1/2), K^*(h = +1) | 
    {\cal L}_{vqq} | \Lambda^* (s_z = 1/2)\ket \, .
    \nonumber
\end{align}
Here $s_z$ represents the third component of spin and $h$ the helicity of 
the photon. In general, for a  massive vector meson, there is another type
of scalar or longitudinal one, which can be computed by the time component 
of the current. For the present purpose, however, the two transverse 
components are sufficient. They are then related to the multipole amplitudes
by 
\begin{align}
    E1 &= - \frac{1}{2} a_{1/2} - \frac{\sqrt{3}}{2} a_{3/2} \, , 
    \nonumber \\
    M2 &= \frac{\sqrt{3}}{2} a_{1/2} - \frac{1}{2} a_{3/2} \, .
    \nonumber
\end{align}
The quark model calculation is rather standard, and so we just show the 
final result:
\begin{align}
    E1 &= -i \frac{3 \sqrt{2} \sqrt{\alpha}}{m_q}
    g_{vqq} \left( 1 - \frac{\bm{k}^2}{6\alpha} \right)
    e^{-\bm{k}^2/6\alpha} \, , 
    \nonumber \\
    M2 &= -i \frac{\sqrt{6} \sqrt{\alpha}}{4m_q}
    g_{vqq} \frac{\bm{k}^2}{6\alpha} e^{-\bm{k}^2/6\alpha} \, , 
    \nonumber
\end{align}
where $\bm{k}$ is the momentum of $K^*$ and $\alpha$ is a harmonic 
oscillator parameter of the wave function of the non-relativistic quark 
model, which is related to the size of the system by 
\ba
\bra r^2 \ket = 3/\alpha \, .
\nonumber
\ea
The $\Lambda^*\bar K^* N$ coupling constant is then related to 
the $E1$ amplitude by an overall constant
\ba
g_{\Lambda^*\bar{K}N} = \frac{3}{\sqrt{6}} E1 .
\ea

\begin{figure}[tbp]
    \centering
    \includegraphics[width=8cm,clip]{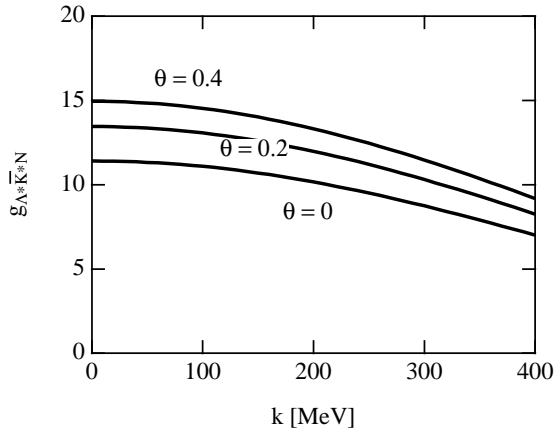}
    \caption{\label{fig:quark}
    Numerical result for the $\Lambda^*\bar{K}^*N$ coupling constant as a 
    function of $K^*$ momentum $k$ in the quark model, for different mixing
    angles $\theta$.}
\end{figure}

In the calculation, we consider a mixing of $^2 \bm{1}$ and $^2 \bm{8}$ 
states for $\Lambda(1520)$ as 
\ba
|\Lambda(1520)\ket = \cos \theta |^2 \bm{1}\ket 
+ \sin \theta |^2 \bm{8}\ket \, .  
\nonumber
\ea
In the Isgur-Karl model, the mixing angle was obtained $\theta \sim 
0.4$~\cite{Isgur:1978xj}. The result is shown in Fig.~\ref{fig:quark}, where
the coupling constant $g_{\Lambda^*\bar{K}^*N}$ is shown as a function of 
$K^*$ three-momentum $k$ for different mixing angles $\theta$. The quark 
model value, in contrast with that of the chiral unitary approach, is of 
order $g_{\Lambda^*\bar{K}^*N} \sim 10$. In particular, the value increases 
slightly as the mixing angle increases, which is a consequence of the 
interference between the two flavor states. The difference between the 
values of the chiral unitary model and the quark model is large, and it 
would be interesting to test the coupling by experiments. In reality, the 
physical resonance state may be a mixture of the two extreme schemes of the 
chiral unitary and the quark models.  The coupling $g_{\Lambda^*\bar{K}^*N}$
could be used to investigate such a hybrid nature of the resonance.  

For completeness, we would like to mention the phenomenological analysis of 
the $g_{\Lambda^*\bar{K}^*N}$ coupling constant. In 
Ref.~\cite{Titov:2005kf}, the $g_{\Lambda^*\bar{K}^*N}$ is estimated from 
the $\Lambda^*$ photoproduction data~\cite{Barber:1980zv}. They fit the 
cross section at $E_{\gamma}=2.8$-$4.8$ GeV by a Regge trajectory of $K^*$ 
exchange, and match the amplitude at $E_{\gamma}=2.3$ GeV to the one 
calculated by the Born terms with the effective Lagrangian approach which 
includes the $g_{\Lambda^*\bar{K}^*N}$ in the $K^*$ $t$-channel exchange. 
The result in the present convention is
\begin{equation}
    g_{\Lambda^*\bar{K}^*N}
    =+7.1 \text{ or } -12.6,
    \label{eq:Titov}
\end{equation}
where we denote the relative $\pm$ sign to $g_{\Lambda^*\bar{K}N}$. However,
this conclusion depends on the assumption of the Regge trajectory of $K^*$ 
exchange, and the same data~\cite{Barber:1980zv} can be equally well 
reproduced with $g_{\Lambda^*\bar{K}^*N}=0$ in a different 
model~\cite{Nam:2005uq}, where the Kroll-Ruderman term plays a dominant 
role. In order to perform a precise phenomenological analysis, we need 
further experimental information of the $\Lambda^*$.

\section{Summary and discussions}

In this paper, we have studied the $\Lambda(1520)\bar{K}^*N$ coupling 
constant. The motivations are twofold: One is to offer a model estimation 
for the unknown coupling constant which is expected to be important in 
hyperon production reactions, and the other one is to test different types 
of models for baryon resonances. In the chiral unitary model the resonances 
are described as a meson baryon quasibound state which may indicate the 
importance of hadron-like correlations in hadron structure.  

Since the coupling constant has not been calculated in the chiral unitary 
model before, we have shown here a detailed derivation. The resulting 
coupling constant $g_{\Lambda^*\bar{K}^*N}$ is expressed as a sum over 
contributions from various channels necessary for the formation of $\Ls$. 
The actual number of the coupling $g_{\Lambda^*\bar{K}^*N}$ turned out to be
of order 1-2, which is significantly smaller than the quark model value of 
order 10. 

The difference in the results in two models should be a consequence of the 
difference of the model setup in various aspects. First, the quark model 
describes the $\Lambda^*$ as a three-quark system, while it is five-quark 
description in the chiral unitary model. Second, in the chiral unitary 
model, the $\Lambda^*$ is mainly a member of flavor $\bm{8}$, while in the 
quark model it is presumably dominated by the flavor singlet $\bm{1}$. 
Third, the wave function of the $\Lambda^*$ would be dominated by the 
$s$-wave component of $\pi\Sigma(1385)$, while it is a $p$-wave excitation 
in the quark model. Such differences in the internal structure should be 
reflected in the $\Lambda^*\bar{K}^*N$ coupling. If the actual $\Lds$ has a 
mixed structure of the meson-baryon quasibound state and the three-quark 
state, the relevant coupling constant will be an intermediate value. 

Since we have no experimental information of the coupling it would be very 
interesting to have the experimental value. Photoproduction reactions such 
as $\gamma N \to \Ls K$ and $\gamma N \to\Ls K^*$ may discriminate the 
coupling constant. In the $K$ production case, comparison between proton 
target and neutron target will be useful, since the $K$ exchange and contact
terms are absent for the neutron target~\cite{Nam:2005uq}. As a consequence,
the $t$-channel behavior is dominated by the $K^*$ exchange, so that the 
angular dependence is very sensitive to the strength of the 
$\Lambda^*\bar{K}^*N$ coupling constant. Hence, the angular dependence of 
the cross section ratio of proton and neutron will give us the information 
of the coupling constant of interest. It is also interesting to investigate 
the $\gamma p \to \Ls K$ and $\gamma p \to\Ls K^*$ reactions with 
$\Lambda^*$ going forward, which is naively dominated by the $u$-channel 
diagram. When the exchanged particle is the $\Lambda^*$, the cross section 
ratio of the $K$ production and the $K^*$ production provides the ratio of 
the coupling constants $\Ls\bar{K}N$ and $\Lambda^*\bar{K}^*N$. Information 
from such experiments as well as theoretical comparison would provide 
further understanding of the resonance structure. 

\begin{acknowledgments}
    One of the authors (T.H.) thanks to the Japan Society for the Promotion
    of Science (JSPS) for support. One of the authors (S.S.) wishes to 
    acknowledge support from the Ministerio de Educacion y Ciencia in the 
    program Doctores y Tecnologos extranjeros. This work is supported in 
    part by the Grant for Scientific Research [(C) No.17959600, T.H.] and 
    [(C) No.16540252, A.H.] from the Ministry of Education, Culture, Science
    and Technology, Japan. This work is partly supported by the contract 
    BFM2003-00856 from MEC (Spain) and FEDER, the Generalitat Valenciana and
    the E.U. EURIDICE network contract HPRN-CT-2002-00311. This research is 
    part of the EU Integrated Infrastructure Initiative Hadron Physics 
    Project under contract number RII3-CT-2004-506078.
\end{acknowledgments}

\appendix
\section{Lagrangians and conventions}

Here we summarize the chiral Lagrangians which are used in the present 
analysis. The coupling of vector meson and pseudoscalar mesons is given by
\begin{equation}
    \cL_1=-i\frac{g}{\sqrt{2}}\Tr(V^\mu[\partial_\mu P,P]),
    \label{eq:vector}
\end{equation}
with $g=-6.05$ and the Yukawa coupling of ground state baryon is given by
\begin{equation}
    \cL_2=\Tr \left(\frac{D}{2}\bar B\gamma^\mu\gamma^5 \{u_\mu, B\}
    + \frac{F}{2}\bar B\gamma^\mu\gamma^5 [u_\mu, B]\right),
    \label{eq:Yukawa}
\end{equation}
with standard notations given in Refs.~\cite{Ecker:1995gg,Pich:1995bw,
Bernard:1995dp}. The coupling constants are such that $D+F=1.26$ and 
$D-F=0.33$. With these Lagrangians, we obtain the amplitudes for the 
$t$-channel meson exchange processes $\bar{K}^*(k) + \pi(q-k)\to \bar{K}(q)$
and $N\to \pi(q-k)+N$
\begin{align}
  -it_1
  =&-i\alpha\frac{g}{2}(2q-k)_\mu\eps^\mu ,
  \nonumber\\
  -it_2
  =&\left(\beta\frac{D+F}{2f}+\gamma\frac{D-F}{2f}\right)
  \bm{\sigma}\cdot(\bm{q}-\bm{k}) ,
  \nonumber
\end{align}
with suitable SU(3) coefficients $\alpha$, $\beta$, and $\gamma$.
The Yukawa coupling of $\Sigma^*\to M_i(q-k)+B_i$
is similarly given by
\begin{equation}
    -it_3=c_i\frac{12}{5}
    \frac{g_A^*}{2f}\bm{S}\cdot(\bm{q}-\bm{k}) .
    \nonumber
\end{equation}
with $g_A^*=(D+F)\times 2.13$ where the numerical factor comes from
$f_{\pi N\Delta}=2.13f_{\pi NN}$. SU(3) coefficients $c_i$ are tabulated in
Refs.~\cite{Jido:2002zk,Hyodo:2004vt}.


\end{document}